# High-pressure structural, elastic, and thermodynamic properties of zircon-type HoPO$_4$ and TmPO$_4$

O. Gomis[1], B. Lavina[2], P. Rodríguez-Hernández[3], A. Muñoz[3], R. Errandonea[4], D. Errandonea[4], and M. Bettinelli[5]

[1]Centro de Tecnologías Físicas, MALTA Consolider Team, Universitat Politècnica de València, 46022 Valencia, Spain

[2]High Pressure Science and Engineering Center and Department of Physics and Astronomy, University of Nevada, Las Vegas, 89154 Las Vegas, Nevada, USA

[3]Departamento de Física, Instituto de Materiales y Nanotecnología, MALTA Consolider Team, Universidad de La Laguna, La Laguna, 38205 Tenerife, Spain

[4]Departamento de Física Aplicada-ICMUV, MALTA Consolider Team, Universidad de Valencia, Edificio de Investigación, C/Dr. Moliner 50, Burjassot, 46100 Valencia, Spain

[5]Laboratory of Solid State Chemistry, DB and INSTM, Università di Verona, Strada Le Grazie 15, 37134 Verona, Italy

E-mail: osgohi@fis.upv.es

**Abstract:** Zircon-type HoPO$_4$ and TmPO$_4$ have been studied by single-crystal x-ray diffraction and *ab initio* calculations. We report information on the influence of pressure on the crystal structure, and on the elastic and thermodynamic properties. The equation of state for both compounds is accurately determined. We have also obtained information on the polyhedral compressibility which is used to explain the anisotropic axial compressibility and the bulk compressibility. Both compounds are ductile and more resistive to volume compression than to shear deformation at all pressures. Furthermore, the elastic anisotropy is enhanced upon compression. Finally, the calculations indicate that the possible causes that make unstable the zircon structure are mechanical instabilities and the softening of a silent $B_{1u}$ mode.

# 1. Introduction

High-pressure (HP) studies of anhydrous rare-earth (RE) orthophosphates (REPO$_4$) are quite important for mineral physics and chemistry. In the last decade such studies have gained attention due to the large number of technological applications of these compounds. REPO$_4$ orthophosphates crystallize either in the tetragonal zircon structure (space group $I4_1/amd$, Z = 4), for the heavy RE-element compounds, or in the monoclinic monazite structure (space group $P2_1/n$, Z = 4) for the light RE-element compounds [1]. Important minerals like Xenotime, in which the major component is YPO$_4$, and Pretulite, in which the major component is ScPO$_4$, belong to the first group. In this work, zircon-type holmium phosphate (HoPO$_4$) and thulium phosphate (TmPO$_4$) have been studied. Their crystal structure (shown in figure 1) is basically made of PO$_4$ tetrahedral units and HoO$_8$ (TmO$_8$) dodecahedra. It is composed of alternating edge-sharing HoO$_8$ (TmO$_8$) and PO$_4$ units, which form chains parallel to the $c$-axis.

REPO$_4$ compounds have been studied under compression by Raman spectroscopy **[2-7]**, x-ray diffraction (XRD) **[4, 6-11]**, and theoretical calculations **[4, 6, 8, 12-14]**. They have also been characterized by inelastic nuclear scattering **[12, 13]** and indentation techniques **[15]**. In addition, the elastic constants of a few REPO$_4$ compounds have been measured (calculated) at ambient pressure (0 GPa) **[12, 16-19]**. Furthermore, the compressibility of the whole series of orthophosphates has been theoretically studied using a chemical-bond theory of dielectric description **[20]**. In previous studies, several pressure-induced structural phase transitions have been discovered. In the case of zircon-structured phosphates, there is an agreement on the fact that depending on the RE metal, the phase transition is either to the tetragonal scheelite structure (space group $I4_1/a$) or to the monazite structure **[4, 6-11]**. In contrast, regarding the compressibility of REPO$_4$ orthophosphates, there is not such agreement.

For instance, in the case of GdPO$_4$, from powder XRD measurements a bulk modulus ($B_0$) of 160(2) GPa is determined **[9]**, while single-crystal XRD experiments led to $B_0$ = 128.1(8) GPa **[7]**. In addition, from theoretical calculations a $B_0$ = 149.0 GPa is determined **[20]**, while from empirical models $B_0$ = 120 GPa is obtained **[21]**. On the other hand, most of the HP studies on REPO$_4$ orthophosphates have been aimed to investigate phase transitions and determine the equation of state (EOS) of different polymorphs. However, little efforts have been devoted for studying the effects of pressure on the elastic properties of orthophosphates.

The facts described above clearly show that more efforts are needed in order to deepen the understanding of the structural and mechanical properties of orthophosphates under compression. With this aim, we have performed a combined theoretical and experimental study of zircon-type HoPO$_4$ and TmPO$_4$. We present HP single-crystal XRD measurements carried out in a diamond-anvil cell (DAC) under quasi-hydrostatic conditions and *ab initio* calculations. We concentrate on discussing the effect of pressure on the structural, mechanical and thermodynamic properties within the pressure range of stability of the zircon-type polymorph. We have found an excellent agreement between experiments and calculations. From calculations, we have determined not only the bulk and polyhedral compressibility but also have obtained the elastic constants, elastic moduli, Poisson's ratio, elastic anisotropy, hardness, sound velocity, Debye temperature, and minimum thermal conductivity. We also discuss the mechanical and dynamical stability of zircon-type HoPO$_4$ and TmPO$_4$.

## 2. Experimental details

Single crystals of HoPO$_4$ and TmPO$_4$ were grown by spontaneous nucleation from a PbO-P$_2$O$_5$ flux (1:1 molar ratio) **[22]**. The reagents employed for the growths were $NH_4H_2PO_4$, PbO (both reagent grade), and $Ho_2O_3$ or $Tm_2O_3$ (99.99%). The batches were put in a covered platinum crucible with a tightly fitting lid and heated up to 1300 °C inside a horizontal furnace. After a soaking time of 15 h, the temperature was decreased to 800 °C with a rate of 1.8 °C/h. The crucible was then drawn out from the furnace and quickly inverted to separate the flux from the crystals grown. The flux was dissolved using hot diluted nitric acid. The crystals obtained were characterized by powder XRD. A single phase of zircon-type $HoPO_4$ and $TmPO_4$ was confirmed, with unit-cell parameters $a$ = 6.8919(8) Å and $c$ = 6.0336(8) Å and $a$ = 6.8219(8) Å and $c$ = 5.9798(8) Å, respectively. The following atomic positions for oxygen (the only atom with symmetry-unconstrained coordinates) were determined for $HoPO_4$ (0, 0.4248(5), 0.2127(5)) and for $TmPO_4$ (0, 0.4234(5), 0.2155(5)). The structural parameters of both compounds are in agreement with earlier reported values [1].

Single-crystal XRD measurements were carried out at room temperature under pressure up to 18 GPa using a piston cylinder and a four-post DACs at 16-IDB beamline of the High Pressure Collaborative Access Team (HPCAT) - Advanced Photon Source (APS) and at beamline 12.2.2 of the Advanced Light Source (ALS). Three independent experiments were performed for $HoPO_4$ and one for $TmPO_4$. Experiments were carried out using monochromatic microfocused Hard X ray (wavelengths of 0.36802, 0.34454 Å and 0.398913 Å). Data were collected with the rotation method with a range of +/- 33 ω degrees; side diffraction images were collected for the unstrained crystal achieving a resolution of 0.55 Å. Data were collected with two area detectors, a MAR345 IP (ALS) and a MARCCD (APS), that were calibrated using the diffraction patterns of $LaB_6$ and $CeO_2$ respectively. The samples used in the experiments were cleaved from the grown single crystals. They were loaded in a 120-μm hole of a rhenium gasket in a DAC with diamond-culet sizes of 300 μm. A ruby sphere and gold powder were also loaded with the sample for pressure determination [23, 24]. Non-hydrostatic stresses have been shown to considerably influence the HP structural behavior of compounds isomorphic to orthophosphates [25-27]; therefore, we used neon as pressure-transmitting medium which has been widely shown to generate quasi-hydrostatic conditions in the pressure range of this work. Detectors calibrations and powder data reduction were performed with FIT2D [28]. Single crystal data reduction and structural refinement were performed with GSE_ADA/ RSV [29] and SHELXL [30].

For $HoPO_4$, high-quality single-crystal XRD patterns were recorded. In the case of $TmPO_4$, the diffraction peaks were broad in $\chi$ but sharp in the $\theta$ direction, indicating a large spread in domains orientation but absence of elastic strain. For this reason, the $TmPO_4$ diffraction data were reduced as powders and only lattice parameters were refined. The unit cell parameters of $HoPO_4$ (single crystal) were calculated through least-squares refinement of 80 to 240 d-spacing values. Structural refinements were performed against squared structure factors after merging for symmetry equivalents ($<R_{eq}>$ = 10%) obtaining satisfactory agreements between observed and calculated intensities ($<R_1>$ = 4.5%). Refined parameters were the scale factor, two oxygen fractional coordinates and the isotropic displacement parameter of Ho; isotropic displacement for O and P were fixed at 0.010 $Å^2$, rounded literature values. When increasing pressure beyond 20 GPa, we found evidence of phase transitions in both compounds, as expected. Since this work is devoted to the properties of the low-pressure phase, the analysis of the results obtained from the post-zircon phases will not be discussed here.

## 3. Theoretical calculation details

The structural and elastic properties of zircon-type $HoPO_4$ and $TmPO_4$ have been studied under pressure performing *ab initio* total energy calculations. The calculations were achieved in the framework of the density functional theory (DFT) **[31]**. It is well stablished that the physical properties of semiconductors under pressure are accurately described by the DFT formalism **[32].**

The Vienna *Ab initio* Simulation Package (VASP) **[33]** was used with the projector-augmented wave scheme (PAW) **[34]**. Thus, the full nodal character of the all-electron charge density in the core region was taken into account. Because the presence of oxygen in the studied compounds, the set of plane waves was developed to an energy cutoff of 520 eV. As for the exchange-correlation energy, it was described within the generalized gradient approximation (GGA) with the Perdew-Burke-Ernzerhof prescription (PBE ) **[35].** The integrations along the Brillouin zone (BZ) were carried out with a dense Monkhorst–Pack grid (6 x 6 x 6) of special k-points. This procedure ensures a high convergence of 1 meV per formula unit in the total energy and accurate values of the forces on the atoms. The lattice parameters and atomic positions of the zircon structure, for both compounds, were fully optimized through the calculation of the forces on the atoms and the stress tensor, at selected volumes. In the resulting optimized structures, the forces on the atoms were less than 0.001 eV/Å, and the deviation of the stress tensor from a diagonal hydrostatic form was lower than 1 Kbar.

In order to study the mechanical properties of zircon-type phosphates, the elastic constants were also determined. The elastic constants were evaluated with the use of the stress theorem **[36]**. The ground state and fully relaxed structures were strained in different directions taking into account their symmetry **[37]**. A Taylor expansion for the total energy with respect to the applied strain was employed to evaluate the total-energy variations **[38]**. The elastic constants describe the mechanical properties of a material in the region of small deformations, so care must be taken that the strain used in the calculations guarantees the harmonic behavior.

Phonons calculations were achieved at the zone center ($\Gamma$ point) of the BZ with the supercell method (or direct force-constant approach) **[39]**. The frequency and symmetry of the phonon modes are provided by the diagonalization of the dynamical matrix. To obtain the phonon dispersion curves along high-symmetry directions of the BZ, we performed similar calculations using appropriate supercells, which allow the phonon dispersion at k-points to be obtained commensurate with the supercell size **[39]**. In this work, we concentrated in the study of the silent $B_{1u}$ mode at $\Gamma$ whose behavior is related to the stability of the zircon structure.

**4. Results and discussion**

**4.1 Effect of pressure on the crystal structure**

First we will compare our calculated crystal structure for both compounds with the experimentally determined crystal structure. According to our calculations, the unit-cell parameters of TmPO$_4$ are *a* = 6.8784 Å and *c* = 5.9922 Å, and the atomic position of oxygen (0, 0.4259, 0.2124). In the case of HoPO$_4$, we obtained *a* = 6.9373 Å and *c* = 6.0391 Å, being the atomic position of oxygen (0, 0.4249, 0.2140). The agreement with the experiments is within 1%, being the unit-cell volumes slightly overestimated by the calculations.

Now we will comment on the effect of HP on the zircon structure. Diffraction images of HoPO$_4$ and TmPO$_4$ collected at HP are shown in figure 2. It can be seen that the pattern of HoPO$_4$ remained that of a nearly unstrained crystal, whereas the TmPO$_4$ peaks were broad in $\chi$ from the beginning and did not appreciably broadened with pressure increase. Therefore, both samples were maintained in quasi-hydrostatic conditions throughout the experiments. The best datasets were collected for HoPO$_4$, experiment 3. The crystal was first measured at ambient conditions in the DAC, obtaining the structural parameters *a*= 6.886(2) Å and *c*= 6.027(2) Å, oxygen coordinates (0, 0.423(4), 0.217(5)), and holmium displacement parameters $U_{eq}$=0.016 (2) Å$^2$. The very satisfactory agreement between the micro-crystal structure refinement and the powder diffraction, as well as the literature [1], validates our structure factors measurements in the diamond anvil cell. The oxygen parameters only slightly changed with pressure reaching the values (0, 0.431(1), 0.216 (2)) at 16.4 GPa. The initial Ho displacement parameter shows a small decrease and then remains constant at around 0.008 Å$^2$.

From our experiments, we found that the zircon phase remains as a single phase up to 17.7 GPa (18.1 GPa) in HoPO$_4$ (TmPO$_4$). Upon further compression, we found in HoPO$_4$ (TmPO$_4$) at 18.8 GPa (21.2 GPa) that reflections corresponding to a different phase coexist with the reflections of the zircon structure. The new phase appears as a single phase at 23.9 GPa in HoPO$_4$ and at 23.4 GPa in TmPO$_4$ and the phase transition is non reversible. Since this work is focused on the behavior of the low-pressure phase, we will only mention here that in both compounds the HP phase has a monoclinic symmetry and unit-cell parameters which indicate that the HP could be isomorphic to the monazite structure. This result is consistent with the results reported previously on TmPO$_4$ and TbPO$_4$ **[4, 8]**.

From the HP experiments, we obtained the pressure dependence of the unit-cell parameters and oxygen coordinates. The same information was obtained from *ab initio* calculations. The results are summarized in figure 3. The agreement between experiments and theory is very good. In the case of TmPO$_4$, the pressure dependence of the lattice parameters is comparable with that obtained from previous powder XRD experiments **[4]**. The ambient pressure results from Ni *et al*. [1] are also included in figure 3 to illustrate the agreement with our results. Regarding the pressure dependence of the structural parameters, experiments and calculations give a quite similar behavior. In particular, we found that the compression of the zircon structure is anisotropic, being the *a*-axis more compressible than the *c*-axis. As a consequence, the *c/a* axial ratio gradually increases under compression (see inset in figure 3). On the other hand, the atomic coordinates of oxygen are slightly modified by pressure.

From the results reported in figure 3, the axial compressibilities of both compounds can be obtained. We have also determined a room-temperature P-V EOS. In our case, the results can be well described by a third-order Birch-Murnaghan (BM) EOS. The axial compressibilities and EOS parameters are summarized in Tables 1 and 2, respectively. The agreement between theory and experiments is excellent.

From the results shown in figure 3, we have determined the pressure dependence of P-O and Ho-O (Tm-O) bond distances as well as the pressure dependence of the volume of different polyhedral units. Since in figure 3 experiments and theory compare quite well, the same conclusions will be extracted from theory than from experiments. Given that calculations show less dispersion than experiments, we have used the calculations to analyze the polyhedral compressibility. In figure 4, we present these results. There, it can be clearly seen than the $PO_4$ tetrahedron is highly uncompressible; however, the $HoO_8$ ($TmO_8$) dodecahedron is much more compressible accounting for most of the volume reduction of the crystal under compression. The pressure dependence of the polyhedral volume can be well described by a third-order BM EOS. The obtained EOS parameters are shown in Table 2. The bulk modulus of the $PO_4$ tetrahedron is comparable to that of ultra-incompressible materials [40, 41]. The bulk modulus of the REO$_8$ dodecahedron is comparable with that of the zircon structure. Consequently, the bulk modulus of the studied compounds can be well described using the model developed by Recio *et al.* [42], to explain the bulk compressibility in spinel-type oxides in terms of the polyhedral compressibility.

Regarding the change induced by pressure in the different polyhedral units, we can state that the PO$_4$ tetrahedron remains as a regular polyhedron with four identical distances within the pressure range of stability of the zircon structure; i.e. there are no evidences of symmetry breaking. In the case of the REO$_8$ dodecahedron, we found the same behavior in the two studied compounds. In these units, there are two short bond distances and two long bond distances. The long bond distance is 2 % longer than the shortest one. The short bonds are oriented mainly perpendicular to the *c*-axis and the long bonds are oriented along it. We found that the long RE-O bonds are less compressible than the short bonds. As a consequence of it, the polyhedral distortion, as defined in VESTA [43], is enhanced by compression increasing from 0.014 at ambient pressure to 0.032 at 20 GPa (see the inset of figure 4). The finding on the differential polyhedral compressibility can be used to explain the anisotropic compressibility of TmPO$_4$ and HoPO$_4$ as previously done for related oxides [44]. In figure 1, it can be seen that along the *c*-axis the REO$_8$ dodecahedra are interconnected by rigid PO$_4$ tetrahedra. However, along the directions perpendicular to the *c*-axis the zircon structure consists of chains of REO$_8$ dodecahedra. This situation makes the *a*-axis to be the direction of easy compression in zircon, making the *a*-axis more compressible than the other axes as found in experiments and calculations.

### 4.2. Elastic properties

The ZrSiO$_4$-type structure of TmPO$_4$ and HoPO$_4$ belongs to the tetragonal Laue group TI. This Laue group contains all crystals with 4*mm*, -42*m*, 422 and 4/*mmm* point groups. In the TI Laue group, there are 6 independent second-order elastic constants **[45]** which, in the Voigt notation, are $C_{11}$, $C_{12}$, $C_{13}$, $C_{33}$, $C_{44}$ and $C_{66}$ **[45, 46]**. When a non-zero uniform stress is applied to the crystal, the elastic properties are described by the elastic stiffness, or stress-strain, coefficients, which are defined as

$$B_{ijkl} = C_{ijkl} + 1/2 \, [\, \delta_{ik}\sigma_{jl} + \delta_{jk}\sigma_{il} + \delta_{il}\sigma_{jk} + \delta_{jl}\sigma_{ik} - 2\,\delta_{kl}\sigma_{ij}\,], \qquad (1)$$

with $C_{ijkl}$ being the elastic constants evaluated at the current stressed state, $\sigma_{ij}$ specify the external stresses, and $\delta_{kl}$ is the Kronecker delta **[47-49]**. In the special case of hydrostatic pressure ($\sigma_{11} = \sigma_{22} = \sigma_{33} = -P$) applied to a tetragonal crystal, the elastic stiffness coefficients in the Voigt notation $B_{ij}$ are: $B_{11} = C_{11} - P$, $B_{12} = C_{12} + P$, $B_{13} = C_{13} + P$, $B_{33} = C_{33} - P$, $B_{44} = C_{44} - P$, and $B_{66} = C_{66} - P$, where $P$ is the hydrostatic pressure. The values of $B_{ij}$ and $C_{ij}$ are equal at 0 GPa. When the elastic stiffness coefficients $B_{ij}$ are used, all relationships of the elasticity theory can be applied for the crystal under any loading, including Born's stability conditions which are identical in both loaded and unloaded states **[48-50]**.

The set of $C_{ij}$ elastic constants calculated at zero pressure for HoPO$_4$ and TmPO$_4$ are shown in **Table 3**. It is found that the six elastic constants are greater in TmPO$_4$ than in HoPO$_4$. Therefore, the stiffness of TmPO$_4$ is greater than that of HoPO$_4$. Our calculated $C_{ij}$ values for both phosphates are in general in good agreement with those calculated in Ref **[12]** shown in **Table 3**. On the other hand, our calculated values of $C_{44}$ and $C_{66}$ in TmPO$_4$ agree reasonable well with the measured values of $C_{44}$ = 67 GPa and $C_{66}$ = 16 GPa for this phosphate (see **Table 3**) [16, 17]. **Table 3** also includes experimental and theoretical elastic constants for other REPO$_4$.

The $C_{33}/C_{11}$ ratio describes the longitudinal elastic anisotropy for a single crystal [51]. This ratio results 1.41 (1.38) for HoPO$_4$ (TmPO$_4$) at 0 GPa, and indicates that the stiffness of HoPO$_4$ (TmPO$_4$) along the *c*-axis is 41% (38%) greater than perpendicular to it, as has been previously discussed. The axial compressibilities $\kappa_a$ and $\kappa_c$ have been obtained using the formulas [52]:

$$\kappa_a = S_{11} + S_{12} + S_{13} \quad (2)$$

$$\kappa_c = 2S_{13} + S_{33} \quad (3)$$

where $S_{ij}$ refers to the components of the elastic compliances tensor. **Table 1** reports the values for $\kappa_a$ and $\kappa_c$, obtained at zero pressure using **Eqs. 2** and **3**, which are in good agreement with those obtained previously from equation of state fits. This result gives us confidence about the correctness of our elastic constants calculations. The degree of elastic anisotropy of a tetragonal single crystal may be also quantified by the ratio between the axial compressibilities, $\kappa_a/\kappa_c$ [53]. The $\kappa_a/\kappa_c$ ratio is 2.83 (2.66) for HoPO$_4$ (TmPO$_4$) at 0 GPa. It is found that $\kappa_a$ is greater than $\kappa_c$ because the *a*-axis is more compressible than the *c*-axis. This is in agreement with the $C_{33}/C_{11}$ ratio greater than 1.

The pressure dependence of the elastic constants, $C_{ij}$, and elastic stiffness coefficients, $B_{ij}$, in HoPO$_4$ and TmPO$_4$ are shown in **figure 5**. All $C_{ij}$ increase with pressure but $C_{66}$. $B_{11}$, $B_{12}$, $B_{13}$ and $B_{33}$ increase with pressure whereas $B_{66}$ decreases with pressure reaching negative values. On the other hand, $B_{44}$ increases up to 3.8 (4.8) GPa in HoPO$_4$ (TmPO$_4$) and decreases at larger pressures.

With the set of six elastic stiffness coefficients, analytical formulas for the bulk ($B$) and shear ($G$) moduli in the Voigt [46], Reuss [54], and Hill [55] approximations, labeled with subscripts $V$, $R$, and $H$, respectively, can be then applied [56]:

$$B_V = \frac{2B_{11} + 2B_{12} + B_{33} + 4B_{13}}{9} \tag{4}$$

$$B_R = \frac{1}{2S_{11} + S_{33} + 2S_{12} + 4S_{13}} \tag{5}$$

$$B_H = \frac{B_V + B_R}{2} \tag{6}$$

$$G_V = \frac{2B_{11} + B_{33} - B_{12} - 2B_{13} + 6B_{44} + 3B_{66}}{15} \tag{7}$$

$$G_R = \frac{15}{8S_{11} + 4S_{33} - 4S_{12} - 8S_{13} + 6S_{44} + 3S_{66}} \tag{8}$$

$$G_H = \frac{G_V + G_R}{2} \tag{9}$$

In the Reuss approximation, we use formulas for $B_R$ and $G_R$ obtained from the elastic compliances $S_{ij}$ tensor. In the Voigt (Reuss) approximation, uniform strain (stress) is assumed throughout the polycrystal [46, 54]. In the Hill approximation, the actual effective $B$ and $G$ elastic moduli are obtained by the arithmetic mean of the Voigt and Reuss values [55]. The Young ($E$) modulus and the Poisson's ratio ($v$) are obtained with the expressions [57]:

$$E_X = \frac{9B_X G_X}{G_X + 3B_X} \tag{10}$$

$$v_X = \frac{1}{2}\left(\frac{3B_X - 2G_X}{3B_X + G_X}\right) \tag{11}$$

where the subscript $X$ refers to the symbols $V$, $R$, and $H$.

**Table 3** reports the elastic moduli at 0 GPa in the Hill approximation for the two phosphates. It is found that the bulk, shear and Young moduli at 0 GPa are larger in $TmPO_4$ than in $HoPO_4$; therefore, the stiffness of $TmPO_4$ is greater than that of $HoPO_4$, as previously commented. The value of the bulk modulus, $B$ = 138.9 GPa (144.1 GPa) in $HoPO_4$ ($TmPO_4$), is in good agreement with experimental values of $B_0$ = 152 GPa (144 GPa) in $HoPO_4$ ($TmPO_4$), and theoretical values of $B_0$ = 146 GPa (142 GPa) in $HoPO_4$ ($TmPO_4$), previously reported in Table 2, which were obtained from fits of experimental and theoretical data to a third-order Birch-Murnaghan equation of state. This agreement is again a validation of the goodness of our calculations of the elastic constants. The results we have obtained for $B_0$ in $TmPO_4$ are in good agreement with the values given in Ref [4], $B_0$ = 135(1) GPa and $B_0$' = 4.7(7) from powder XRD experiments, and $B_0$ = 140 GPa and $B_0$' = 5.3 from *ab initio* calculations. The values of $B_0$ = 143.4 GPa ($HoPO_4$) and $B_0$ = 147.2 GPa ($TmPO_4$) obtained using the chemical bond theory of dielectric description [20], and $B_0$ = 142.3 GPa ($HoPO_4$) and $B_0$ = 146.2 GPa ($TmPO_4$) using an empirical model [21], are in nice agreement to those reported in our work.

**Table 3** also includes the Poisson's ratio (*v*), the ratio between the bulk and shear modulus (*B/G*), and the universal elastic anisotropy index ($A_U$) at 0 GPa. The Poisson's ratio provides information about the characteristics of the bonding forces and chemical bonding. Both phosphates have similar values of the Poisson's ratio in the Hill approximation: *v* = 0.30 (0.29) in $HoPO_4$ ($TmPO_4$). This value shows that the interatomic bonding forces are predominantly central (*v* > 0.25) and that ionic bonding is predominant against covalent bonding at 0 GPa **[58, 59].**

Pugh **[60]** has proposed a simple relationship, empirically linking the plastic properties of materials with their elastic moduli. The shear modulus *G* represents the resistance to plastic deformation, while the bulk modulus $B$ represents the resistance to fracture. A high $B/G$ ratio may then be associated with ductility whereas a low value would correspond to a more brittle nature. The critical value which separates ductile and brittle materials is around 1.75, i.e. if B/G > 1.75 the material behaves in a ductile manner; otherwise the material behaves in a brittle manner. In our study, we have found values of $B/G$ at 0 GPa in the Hill approximation above 1.75 for HoPO$_4$ and TmPO$_4$. Therefore, both compounds are ductile at zero pressure, being HoPO$_4$ more ductile than TmPO$_4$.

One of the elastic properties of crystals with more importance for both engineering science and crystal physics is the elastic anisotropy, because it is highly correlated to the possibility of inducing microcraks in the materials **[61].** This anisotropy can be quantified with the universal elastic anisotropy index, $A_U$, introduced by Ranganathan *et al*. **[62]**, which is defined as $A_U=5(G_V/G_R)+(B_V/B_R)-6$, where $B_V$, $G_V$, $B_R$ and $G_R$ are the bulk and shear moduli in the Voigt and Reuss approximations, respectively. We note that $A_U$ is applicable to all types of crystals (symmetries) and takes into account all the stiffness coefficients $B_{ij}$ by recognizing the tensorial nature of this physical magnitude **[62]**. If $A_U$ is equal to 0, no anisotropy exists. On the other hand, the more this parameter differs from 0 the more elastically anisotropic is the crystalline structure. The $A_U$ values for the phosphates are above 0 at zero pressure; therefore, they are anisotropic, being the anisotropy of HoPO$_4$ slightly greater than that of TmPO$_4$. The elastic anisotropy of both phosphates reported by $A_U$ is in agreement with the longitudinal elastic anisotropy given by the $C_{33}/C_{11}$ ratio and the anisotropy in the axial compressibilities given by the $\kappa_a/\kappa_c$ ratio, both previously commented.

**Figure 6** shows the pressure dependence of $B$, $G$, and $E$ elastic moduli for HoPO$_4$ and TmPO$_4$ in the Hill approximation. It can be seen that the bulk modulus increases with increasing pressure. Contrarily, the *G* and *E* moduli decrease with increasing pressure. On the other hand, the two phosphates are more resistive to volume compression than to shear deformation ($B > G$) at any pressure.

**Figure 7(a), 7b) and 7(c)** show the Poisson's ratio, ν, *B/G* ratio, and $A_U$ as a function of pressure. The Poisson's ratio increases with pressure, reaching a value of 0.40 (0.40) at 12.0 GPa (14.0 GPa) for $HoPO_4$ ($TmPO_4$). This indicates an increment of the ductility and a progressive loss of the ionic character with increasing pressure. Similarly, the *B/G* ratio is related to the Poisson's ratio **[59]** and also increases with pressure in the two phosphates, thus indicating an increment of the ductility with pressure. The *B/G* ratio reaches a value of 4.54 (4.46) at 12.0 GPa (14.0 GPa) in $HoPO_4$ ($TmPO_4$). Finally, the $A_U$ universal anisotropy factor increases with increasing pressure in $HoPO_4$ and $TmPO_4$. Therefore, the elastic anisotropy increases in both compounds as pressure increases.

Hardness is generally related to the elastic and plastic properties of a material. The Vickers hardness, $H_V$, can be obtained by the equation proposed by Tian *et al.* **[63]**:

$$H_V = 0.92(G/B)^{1.137} G^{0.708} \qquad (12)$$

We highlight that hardness formula given by Equation (12) is an approximate equation obtained from a fitting to a dataset. In this way, hardness is approximated by a model. On the other hand, we used formula given in Ref **[63]** as it eliminates the possibility of unrealistic negative hardness. **Table 3** includes the values of $H_V$ for $HoPO_4$ and $TmPO_4$ at 0 GPa in the Hill approximation. $TmPO_4$ is harder than $HoPO_4$ and both have values of $H_V$ of approximately 7-8 GPa (~750 Vickers hardness). Taking into account the values of $H_V$, both compounds can be classified as relatively soft materials. The soft behavior of both phosphates is correlated with their predicted ductility at zero pressure as previously shown. These values of hardness are higher than those measured by indentation for zircon-type DyPO$_4$ and (Gd$_{0.4}$Dy$_{0.6}$)PO$_4$, (Gd$_{0.6}$Dy$_{0.4}$)PO$_4$, and (Gd$_{0.5}$Dy$_{0.5}$)PO$_4$ which have hardnesses in the range 3.2 to 6.1 GPa **[15]**.

**Figure 7(d)** shows the pressure evolution of the Vickers hardness with pressure. It can be seen that $H_V$ decreases as pressure increases for both phosphates. This is because the $G/B$ ratio and the $G$ elastic modulus decrease with pressure. In this way, as pressure increases, both phosphates become softer in good agreement with the increase of their ductility ($B/G$ ratio) as stated above.

Finally, one elastic property which is fundamental for Earth Sciences in order to interpret seismic waves is the average sound velocity, $v_m$ **[64]**. In polycrystalline materials $v_m$ is given by **[65]**:

$$v_m = \left[\frac{1}{3}\left(\frac{2}{v_{trans}^3} + \frac{1}{v_{lon}^3}\right)\right]^{-1/3} \quad (13)$$

where $v_{trans}$ and $v_{lon}$ are the transverse and longitudinal elastic wave velocities of the polycrystalline material which are given by:

$$v_{lon} = \left(\frac{B + \frac{4}{3}G}{\rho}\right)^{1/2} \quad (14)$$

$$v_{trans} = \left(\frac{G}{\rho}\right)^{1/2} \quad (15)$$

where $B$ and $G$ are the elastic moduli and $\rho$ the density. **Table 4** includes values of the density, and wave velocities $v_m$, $v_{lon}$ and $v_{trans}$ at 0 GPa for the two phosphates. Wave velocities are slightly greater for TmPO$_4$ than for HoPO$_4$.

**Figure 8** reports the evolution of the elastic wave velocities for both phosphates in the Hill approximation. The calculated $v_{lon}$ increases with pressure reaching a maximum value of 6312.7 m/s (6370.6 m/s) at 8.0 GPa (9.9 GPa) for HoPO$_4$ (TmPO$_4$) and decreases above that pressure. On the other hand, the velocities $v_{trans}$ and $v_m$ decrease as pressure increases.

To conclude this section, we highlight that it is interesting the study and visualization of the elastic properties in any direction. A complete analysis and representation of anisotropic elastic properties such as Young's modulus, shear modulus, linear compressibility, Poisson's ratio, and sound velocity can be carried out with the ElAM (Elastic Anisotropy Measures) computer program **[66]**. However, this study is beyond the scope of the present work and merits an independent research.

### 4.3. Thermodynamic properties

The Debye temperature is an important parameter that correlates with many physical properties of solids, such as specific heat, elastic constants, and melting temperature. One of the standard methods of calculating the Debye temperature, $\theta_D$, is from elastic constant data using the semi-empirical formula **[65]**:

$$\theta_D = \frac{h}{k_B}\left[\frac{3n}{4\pi}\left(\frac{N_A \rho}{M}\right)\right]^{1/3} v_m \qquad (16)$$

where $h$ is the Planck's constant, $k_B$ is the Boltzmann's constant, $n$ is the number of atoms in the molecule, $N_A$ is the Avogadro's number, $\rho$ is the density, $M$ is the molecular weight, and $v_m$ is the average sound velocity. The values of $\theta_D$ at 0 GPa using the Hill approximation, given in **Table 4**, are 475.0 K (486.8 K) in HoPO$_4$ (TmPO$_4$). **Figure 9(a)** shows the evolution with pressure of the Debye temperature, $\theta_D$, for both phosphates. It is found that $\theta_D$ decreases with pressure because $v_m$ decreases with pressure.

The thermal conductivity is the property of a material that indicates its ability to conduct heat. In other to estimate the theoretical minimum thermal conductivity, the following expression has been used [67]:

$$\kappa_{min} = k_B v_m \left( \frac{M}{n\rho N_A} \right)^{-2/3} \quad (17)$$

The values of $\kappa_{min}$ at 0 GPa in HoPO$_4$ (TmPO$_4$) using the Hill approximation are 0.96 (0.99) W m$^{-1}$ K$^{-1}$ (see **Table 4**). This indicates that both phosphates are low $\kappa$ materials [68]. As expected, the values obtained for the minimum thermal conductivity in HoPO$_4$ and TmPO$_4$ are smaller (lower bound) than the values measured for the thermal conductivity, $\kappa$, in other zircon-type REPO$_4$ (YPO$_4$, ErPO$_4$, YbPO$_4$, and LuPO$_4$) with $\kappa \sim 12$ W m$^{-1}$ K$^{-1}$ [69].

**Figure 9(b)** reports the evolution with pressure of the minimum thermal conductivity, $\kappa_{min}$, for both phosphates. As in the case of $\theta_D$, $\kappa_{min}$ decreases with pressure because of the decreasing of $v_m$ with pressure.

### 4.4. Mechanical and dynamical stability of the crystalline structure

The mechanical stability of a crystal at zero pressure can be studied with the Born stability criteria [70]. The conditions for elastic stability at a given pressure $P$, known as the generalized stability criteria, are given for a tetragonal crystal by [71-72]:

$$M_1 = B_{11} > 0, \quad (18)$$

$$M_2 = B_{11} - B_{12} > 0 \quad (19)$$

$$M_3 = (B_{11} + B_{12})B_{33} - 2B_{13}^2 > 0 \quad (20)$$

$$M_4 = B_{44} > 0 \qquad (21)$$

$$M_5 = B_{66} > 0, \qquad (22)$$

**Figure 10** shows the pressure dependence of the generalized stability criteria for HoPO$_4$ and TmPO$_4$. Our calculations show that the five criteria are satisfied for the two phosphates at 0 GPa, thus the zircon-type structure is mechanically stable at 0 GPa, as expected. In HoPO$_4$ (TmPO$_4$) all stability criteria are satisfied at high pressure except $M_5$ which is violated at 12.6 GPa (14.7 GPa). We highlight that the condition that first is violated (Eq. (22)) is related with a pure shear instability because of the decreasing of $B_{66}$ with pressure. On the other hand, we note that the $A_U$ universal anisotropy factor increases quickly at high pressure when the compound approaches the mechanical instability (see **figure 7(c)**).

It has been earlier proposed for TmPO$_4$, a connection between the softening of a silent $B_{1u}$ mode and the fact that the zircon structure becomes unstable **[4]**. A similar correlation was reported in TbVO$_4$ **[73]**. This mode has been assigned to rotations of rigid (PO$_4$)$^{3-}$ units **[74]**. In our case, we have calculated the pressure dependence of this mode for the two studied compounds. We confirm the previous results for TmPO$_4$ **[4]** and found a qualitative similar behaviour in HoPO$_4$. The results of the present calculations are shown in figure 11. In both compounds the frequency of the $B_{1u}$ mode of the zircon structure becomes imaginary near 17 GPa, which is consistent with the fact that near 20 GPa the zircon structure becomes unstable. However, the fact that the $M_5$ generalized stability criterion is violated at pressure lower than 17 GPa (see above) indicates that, in spite that both dynamical and mechanical instabilities could trigger a phase transition under compression in TmPO$_4$ and HoPO$_4$, the main factor activating the phase transition in zircon-type oxides under compression could be mechanical instabilities. This fact is consistent with the plastic deformation reported for zircon-type GdVO$_4$ under compression **[75].** To conclude, we have also calculated the phonon dispersion curves along the whole BZ in the pressure range 0-20 GPa. We only found dynamical instabilities around the BZ center at pressures where the $B_{1u}$ mode becomes also dynamically unstable.

## 5. Conclusions

We have studied experimentally and theoretically the pressure effects on the crystal structure of HoPO$_4$ and TmPO$_4$. This is the first time that single-crystal XRD experiments under quasi-hydrostatic conditions have been carried out in both compounds. The equation of state and the axial and polyhedral compressibility are determined for them. The agreement between experiments and calculations is excellent. We have also theoretically studied the elastic and thermodynamic behavior of the two zircon-type phosphates at high pressure. It has been found that TmPO$_4$ is stiffer than HoPO$_4$. Both compounds are ductile and more resistive to volume compression than to shear deformation ($B > G$) at all pressures. Furthermore, the elastic anisotropy increases with increasing pressure in both phosphates. The two compounds are relatively soft at 0 GPa and their hardness decrease with increasing pressure. The average elastic wave velocity, Debye temperature and minimum thermal conductivity of both phosphates decreases with increasing pressure and are lower in HoPO$_4$ than in TmPO$_4$. We have studied the mechanical stability of the zircon-type structure at high pressure in both compounds, and have found that this structure becomes mechanically unstable at 12.6 GPa (14.7 GPa) in HoPO$_4$ (TmPO$_4$). Finally, we also found that in the two studied compounds there is a silent soft mode which becomes imaginary near the transition pressure. Thus, apparently the zircon structure becomes unstable under pressure due to mechanical and dynamical instabilities. The present results will contribute to improve the understanding of the high-pressure behavior of zircon-type oxides as similar studies did for related monazite-type oxides **[76-78].**

**Acknowledgements**

This research is partially supported by the Spanish government MINECO under Grants No: MAT2016-75586-C4-1-P/2-P/3-P, MAT2013-46649-C4-1-P/2-P/3-P and MAT2015-71070-REDC. A.M. and P.R-H. acknowledge computing time provided by Red Española de Supercomputación (RES) and MALTA-Cluster. This work was also partially supported by the National Nuclear Security Administration under the Stewardship Science Academic Alliances program through DOE Cooperative Agreement No. DE-NA0001982. Part of this work was conducted at HPCAT (Sector 16), Advanced Photon Source (APS), Argonne National Laboratory and at the Advanced Light Source (ALS), Lawrence Berkeley National Laboratory (LBNL). HPCAT operations are supported by DOE- NNSA under Award No. DE-NA0001974 and DOE-BES under Award No. DE-FG02-99ER45775, with partial instrumentation funding by NSF. APS is supported by DOE-BES, under Contract No. DE-AC02-06CH11357. The Advanced Light Source is supported by the Director, Office of Science, Office of Basic Energy Sciences of the U.S. Department of Energy under Contract No. DE-AC02-05CH11231. Use of the COMPRES-GSECARS gas loading system was supported by COMPRES under NSF Cooperative Agreement EAR 11-57758 and by GSECARS through NSF grant EAR-1128799 and DOE grant DE-FG02-94ER14466. We thank Yue Meng, Jesse Smith, Bora Kalkhan and Sergey Tkachev for their assistance.

**Table 1.** Experimental (exp) and theoretical (theory) values for $\kappa_a$ and $\kappa_c$ axial compressibilities in HoPO$_4$ and TmPO$_4$ along with the $\kappa_a/\kappa_c$ ratio at 0 GPa.

| Compound | $\kappa_a$ (10$^{-3}$ GPa$^{-1}$) | $\kappa_c$ (10$^{-3}$ GPa$^{-1}$) | $\kappa_a/\kappa_c$ | |
|---|---|---|---|---|
| HoPO$_4$ | 3.18 | 1.12 | 2.83 | theory[a] |
| HoPO$_4$ | 3.15(3) | 1.16(1) | 2.70(5) | theory[b] |
| HoPO$_4$ | 2.9(1) | 1.2(2) | 2.4(4) | exp[c] |
| TmPO$_4$ | 3.02 | 1.14 | 2.66 | theory[a] |
| TmPO$_4$ | 3.00(2) | 1.13(2) | 2.66(7) | theory[b] |
| TmPO$_4$ | 3.0(5) | 1.0(2) | 2.9(9) | exp[c] |

[a] Obtained from the elastic constants.

[b] Obtained from a Murnaghan equation of state fit to theoretical data.

[c] Obtained from a Murnaghan equation of state fit to experimental data.

**Table 2:** EOS determined for the studied compounds. The EOS for the polyhedral units is also reported. The volume ($V_0$), bulk modulus ($B_0$), its pressure derivative ($B_0'$), and the implied value of the second pressure derivative ($B_0''$) are given.

|  | $V_0$ (Å³) | $B_0$ (GPa) | $B_0'$ (dimensionless) | $B_0''$ (GPa⁻¹) |
|---|---|---|---|---|
| HoPO₄ (exp) | 285.9(7) | 152(3) | 4.20(9) | -0.0273 |
| HoPO₄ (theory) | 290.6(1) | 146(1) | 4.67(4) | -0.0343 |
| PO₄ (theory) | 1.91(2) | 341(4) | 6.28(5) | -0.0331 |
| HoO₈ (theory) | 23.3(1) | 126(2) | 4.45(9) | -0.0359 |
| TmPO₄ (exp) | 281.1(7) | 144(3) | 3.95(9) | -0.0267 |
| TmPO₄ (theory) | 283.5(1) | 142(1) | 5.06(4) | 0.0428 |
| PO₄ (theory) | 1.90(2) | 338(4) | 6.58(3) | -0.0385 |
| TmO₈ (theory) | 22.5(1) | 121(2) | 4.76(9) | -0.0432 |

**Table 3.** $C_{ij}$ elastic constants (in GPa) for HoPO$_4$ and TmPO$_4$. Elastic moduli *B*, *G*, and *E* (in GPa), Possion's ratio (*v*), *B/G* ratio, and Vickers Hardness, $H_V$ (in GPa), are given in the Hill approximation. The universal anisotropy index ($A_U$) is also included. All theoretical data are reported at 0 GPa. In addition, calculated and experimental elastic constants for some REPO$_4$ already reported **[12, 16-19]** are given for comparison. In this latter case, the table also includes the magnitudes than can be obtained from the elastic constants using the formula given in this text.

|       | HoPO$_4$[a] | TmPO$_4$[a] | HoPO$_4$[b] | TmPO$_4$[b] | TmPO$_4$[c] | YPO$_4$[d] | YbPO$_4$[e] | LuPO$_4$[f] |
|-------|-------------|-------------|-------------|-------------|-------------|------------|-------------|-------------|
| $C_{11}$ | 253.6 | 265.7 | 265 | 275 | -- | 220 | 292 | 320 |
| $C_{12}$ | 27.8 | 28.8 | 36 | 40 | -- | 55 | 22 | 36 |
| $C_{13}$ | 94.4 | 96.3 | 130 | 136 | -- | 86 | -- | 115 |
| $C_{33}$ | 356.4 | 367.1 | 401 | 411 | -- | 332 | 315 | 382 |
| $C_{44}$ | 72.8 | 78.5 | 77 | 81 | 67 | 64.6 | 87 | 84.6 |
| $C_{66}$ | 19.9 | 22.5 | 7 | 10 | 16 | 17.3 | 35 | 21.7 |
| *B* | 138.9 | 144.1 | 159.1 | 166.0 | | 132.4 | | 169.3 |
| *G* | 63.9 | 68.8 | 50.5 | 56.1 | | 55.2 | | 73.3 |
| *E* | 166.2 | 178.1 | 137.0 | 151.3 | | 145.5 | | 192.1 |
| *v* | 0.30 | 0.29 | 0.36 | 0.35 | | 0.32 | | 0.31 |
| *B/G* | 2.17 | 2.09 | 3.15 | 2.96 | | 2.40 | | 2.31 |
| $A_U$ | 2.47 | 2.23 | 9.24 | 6.36 | | 2.38 | | 2.68 |
| $H_V$ | 7.22 | 7.94 | 4.00 | 4.64 | | 5.83 | | 7.42 |

[a] This work (theoretical calculations).

[b] Reference **[12]** (theoretical calculations).

[c] Reference **[16]** (experimental data).

[d] Reference **[17]** (experimental data).

[e] Reference **[18]** (experimental data).

[f] Reference **[19]** (experimental data

**Table 4.** Longitudinal ($v_{lon}$ in m/s), transverse ($v_{trans}$ in m/s) and average ($v_m$ in m/s) elastic wave velocity, Debye temperature ($\theta_D$ in K), and minimum thermal conductivity ($\kappa_{min}$ in W m$^{-1}$ K$^{-1}$) in HoPO$_4$ and TmPO$_4$. Data are reported in the Hill approximation at 0 GPa. The density ($\rho$ in g/cm$^3$) is also included. The table also includes the magnitudes than can be obtained from the elastic constants reported in Refs **[12, 17, 19]** using the formula given in this text along with $\rho$.

|  | HoPO$_4$ [a] | TmPO$_4$ [a] | HoPO$_4$ [b] | TmPO$_4$ [b] | YPO$_4$ [c] | LuPO$_4$ [d] |
|---|---|---|---|---|---|---|
| $v_{lon}$ | 6141.9 | 6176.8 | 6170.2 | 6239.3 | 6650.3 | 6395.8 |
| $v_{trans}$ | 3279.7 | 3336.6 | 2913.2 | 3011.7 | 3443.0 | 3350.6 |
| $v_m$ | 3663.6 | 3724.1 | 3278.2 | 3385.3 | 3854.1 | 3747.7 |
| $\theta_D$ | 475.0 | 486.8 | 425.2 | 442.6 | 501.6 | 495.1 |
| $\kappa_{min}$ | 0.96 | 0.99 | 0.86 | 0.90 | 1.02 | 1.02 |
| $\rho$ | 5.940 | 6.183 | 5.948 | 6.187 | 4.659 | 6.528 |

[a] This work.

[b] Reference **[12]**.

[c] Reference **[17].**

[d] Reference **[19].**

**Figure captions**

**Figure 1. (color online)** Two different views of the crystal structure of zircon-type orthophosphates. The $PO_4$ tetrahedra and $REO_8$ dodecahedra are shown.

**Figure 2.** Diffraction patterns of $HoPO_4$ (left) and $TmPO_4$ (right) at 16.4 and 14.5 GPa respectively. The strongest diffraction peaks are generated by the diamond anvils; the broader single crystal peaks in the left image are Neon peaks. Few labels and $HoPO_4$ indices are reported in the zoomed image in the inset.

**Figure 3. (color online)** (Top) pressure dependence of the unit-cell parameters. Different symbols correspond to different experiments. In the case of $TmPO_4$, results from previous powder XRD experiments from Stavrou *et al.* **[4]** are included. The red stars are the ambient pressure results from Ni et al. **[1]**. Solid lines are the results of present *ab initio* calculations. The inset shows the $c/a$ ratio. (Bottom) pressure dependence of the unit-cell volume. For the symbols, we used the same nomenclature that in the top panels. The solid lines are the results of calculations and the dashed lines the results of the EOS fit described in the text. The inset shows the pressure dependence of the oxygen atomic coordinates including our results for $HoPO_4$ and previous results from Stavrou *et al.* for $TmPO_4$ **[4]** (symbols) and present calculations (lines).

**Figure 4.** Relative change of the bulk and polyhedral volumes. The inset shows the pressure dependence of the distortion index of the $HoO_8$ (solid line) and $TmO_8$ (dashed line) dodecahedron.

**Figure 5. (Color online)** Pressure dependence of the theoretical elastic constants ($C_{ij}$) and elastic stiffness coefficients ($B_{ij}$) in $HoPO_4$ (solid lines) and $TmPO_4$ (dashed lines).

**Figure 6**. **(Color online)** Pressure dependence of the elastic moduli $B$, $G$, and $E$ in HoPO$_4$ (solid lines) and TmPO$_4$ (dashed lines). Data are reported in the Hill approximation.

**Figure 7.** Pressure dependence of the Poisson's ratio ($\nu$), $B/G$ ratio, Universal anisotropy index ($A_U$), and Vickers hardness ($H_V$) in HoPO$_4$ (solid lines) and TmPO$_4$ (dashed lines). $\nu$, $B/G$, and $H_V$ are reported in the Hill approximation.

**Figure 8. (Color online)** Pressure dependence of the longitudinal ($v_{lon}$), transverse ($v_{trans}$), and average ($v_m$) elastic wave velocity in HoPO$_4$ (solid lines) and TmPO$_4$ (dashed lines). Data are reported in the Hill approximation.

**Figure 9.** Evolution with pressure of the Debye temperature ($\theta_D$), and the minimum thermal conductivity ($\kappa_{min}$) in HoPO$_4$ (solid lines) and TmPO$_4$ (dashed lines). Data are reported in the Hill approximation.

**Figure 10.** (**Color online**) General stability criteria in HoPO$_4$ (solid lines) and TmPO$_4$ (dashed lines). The pressure at which each phosphate becomes mechanically unstable is shown.

**Figure 11. (Color online)** Calculated frequency of the $B_{1u}$ silent mode of the zircon-type phase of TmPO$_4$ and HoPO$_4$ as a function of pressure.

**Figure 1**

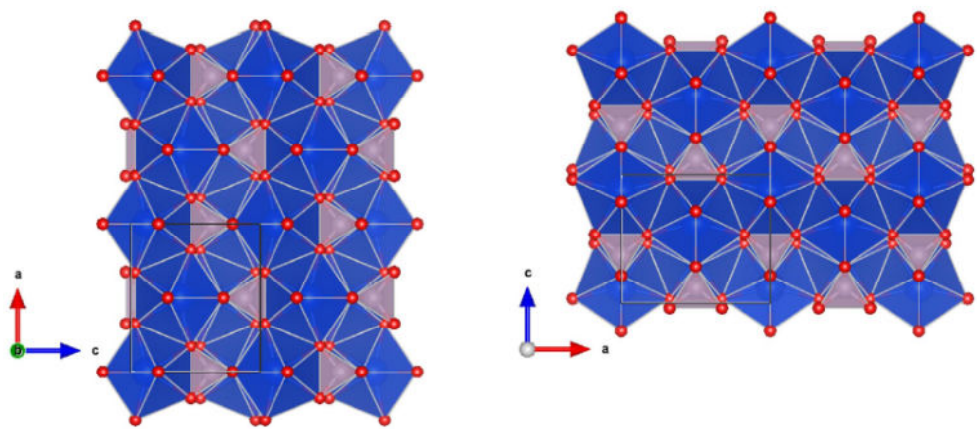

**Figure 2**

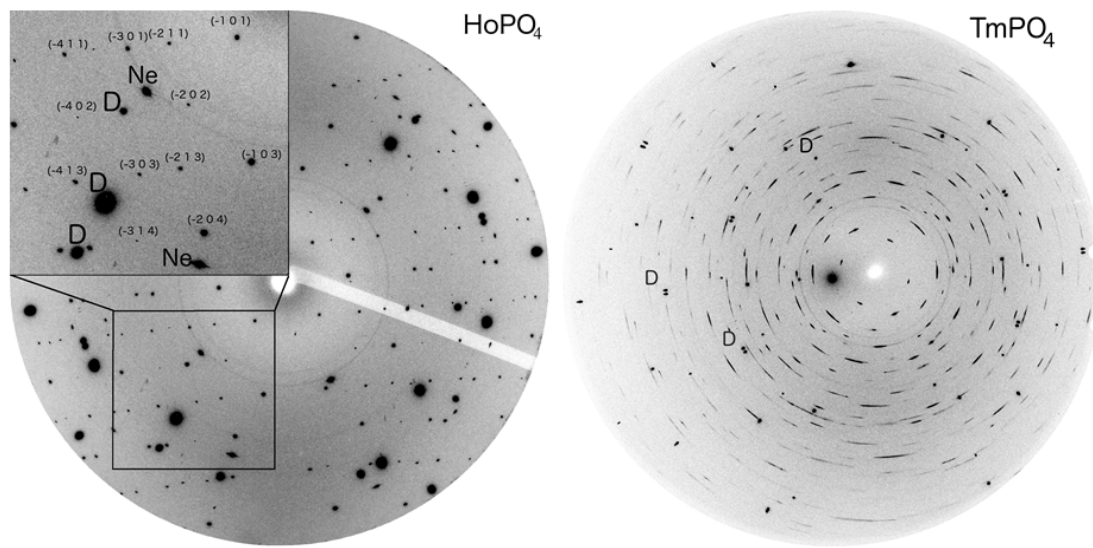

**Figure 3**

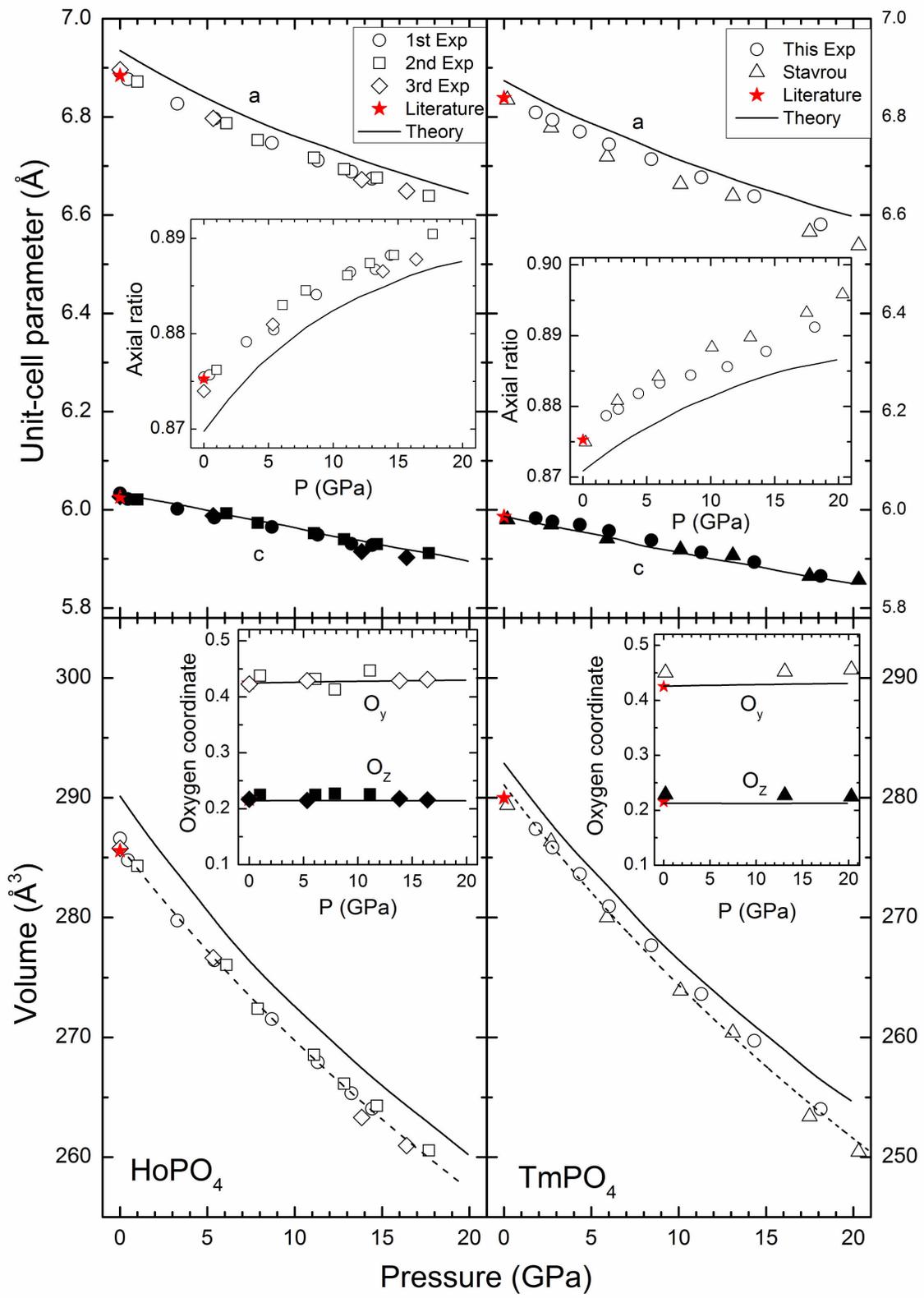

**Figure 4**

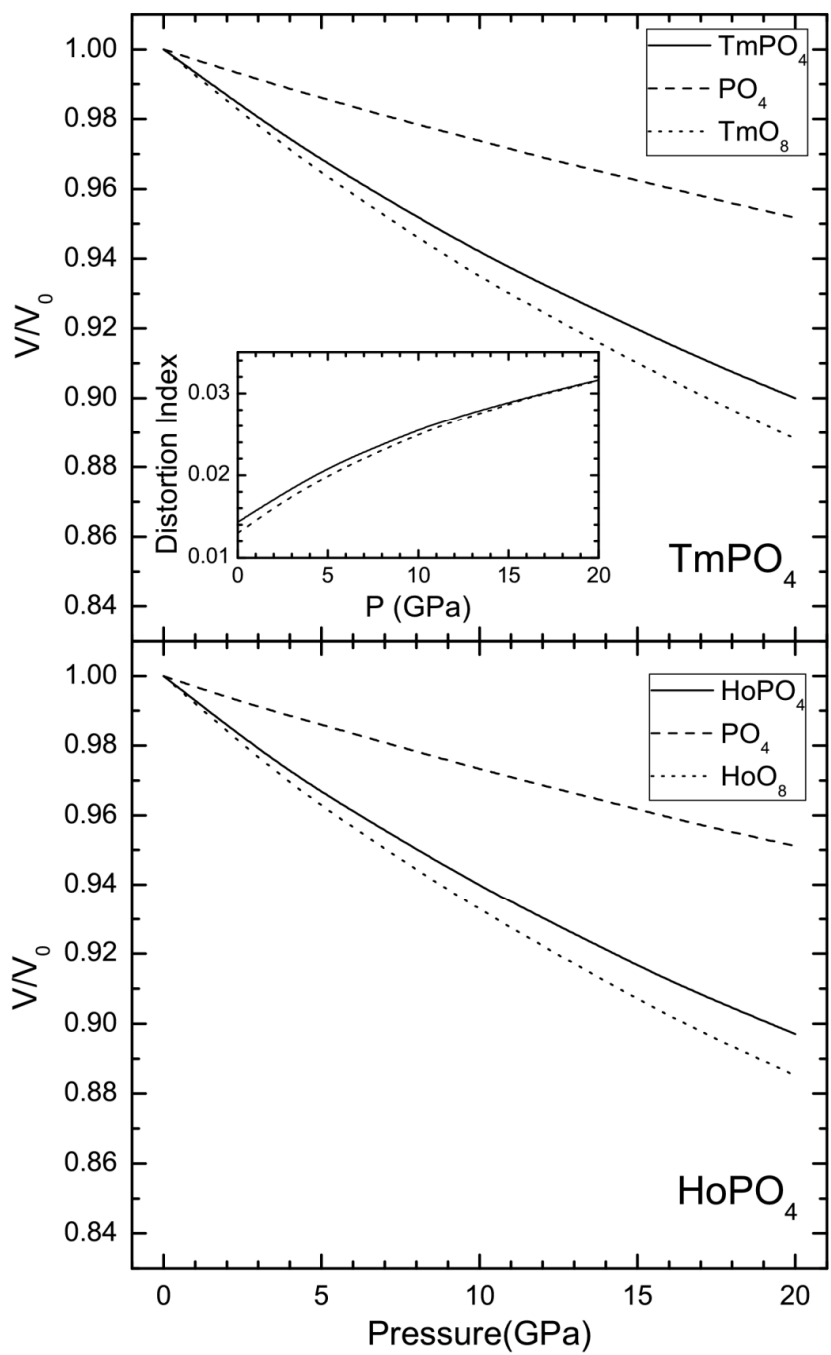

**Figure 5**

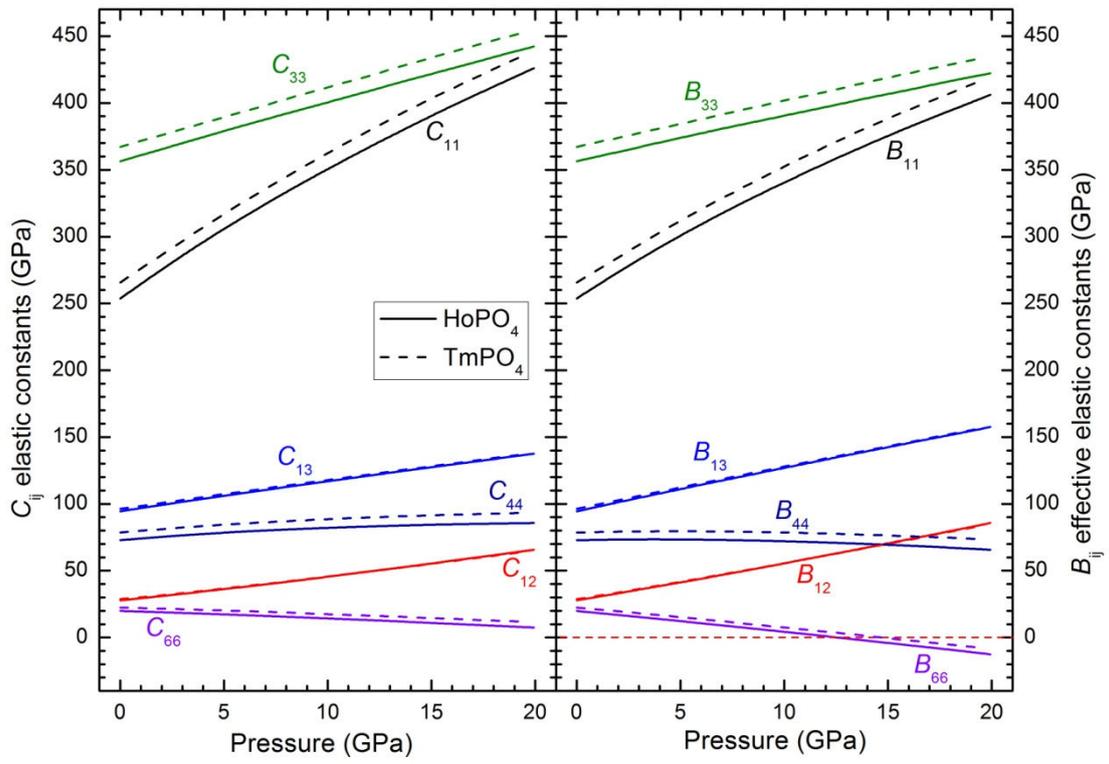

**Figure 6**

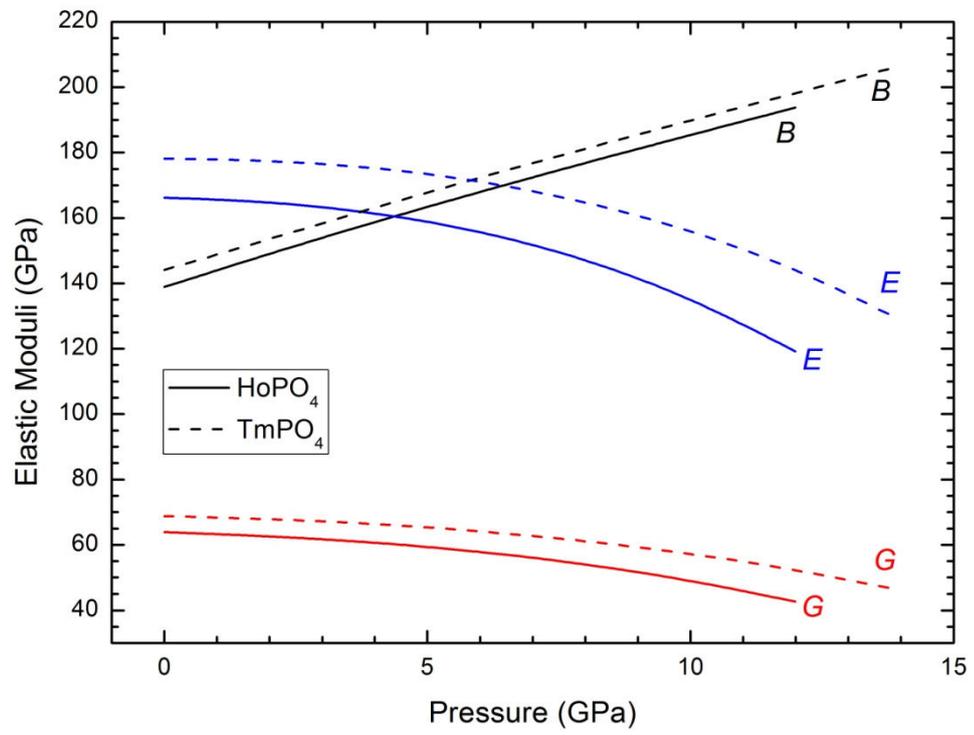

**Figure 7**

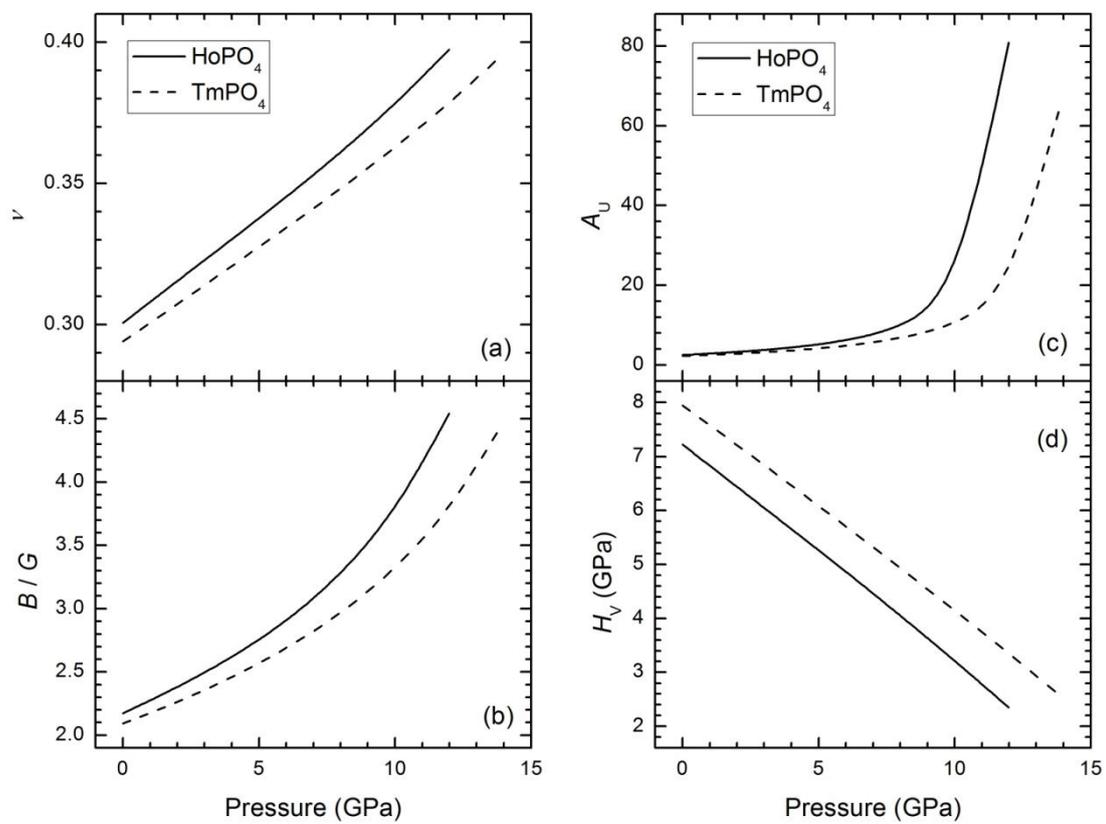

**Figure 8**

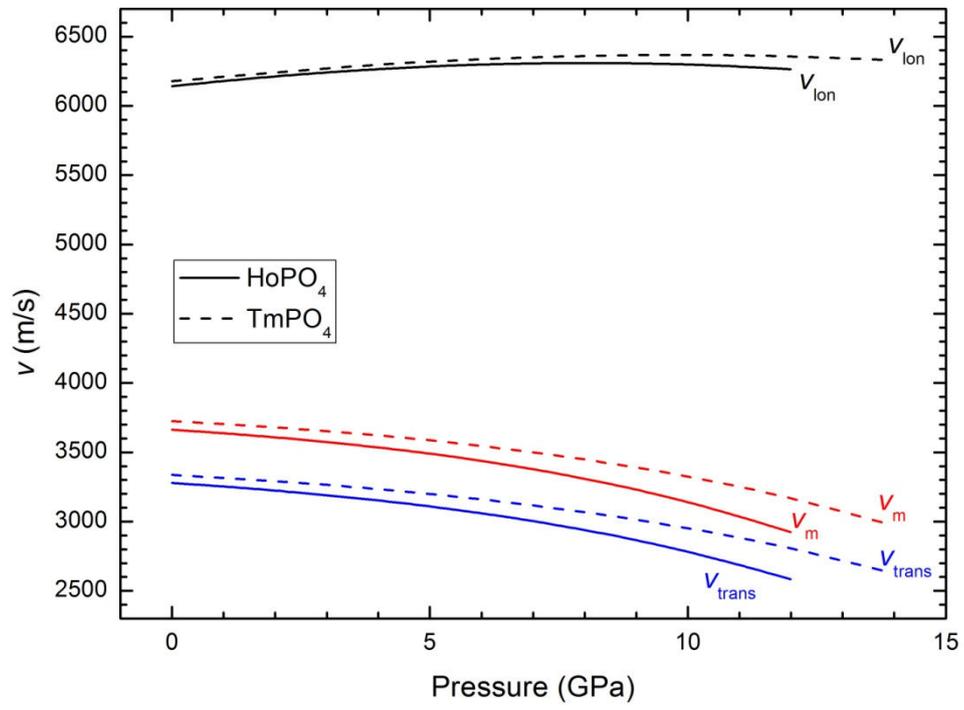

**Figure 9**

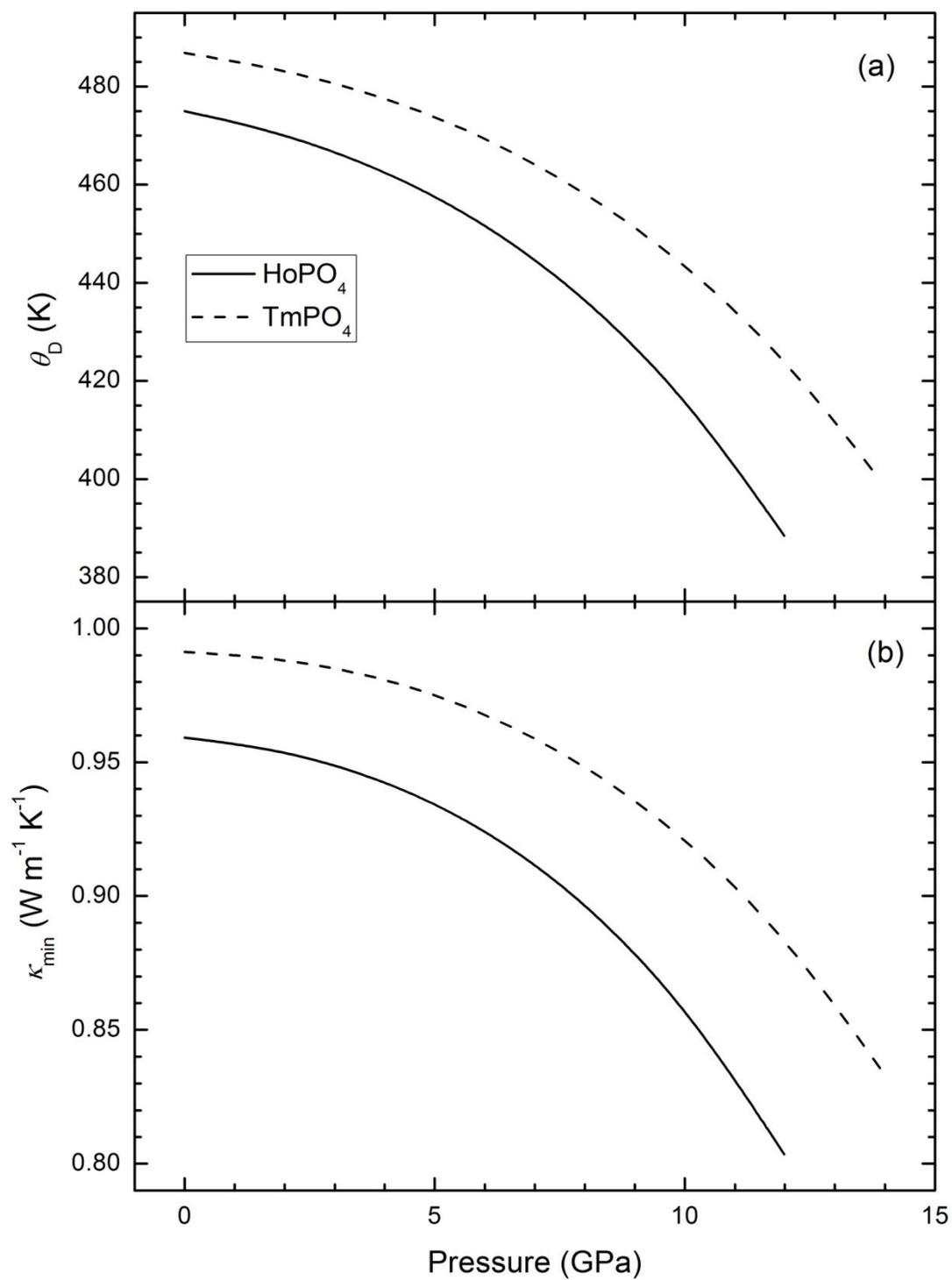

**Figure 10**

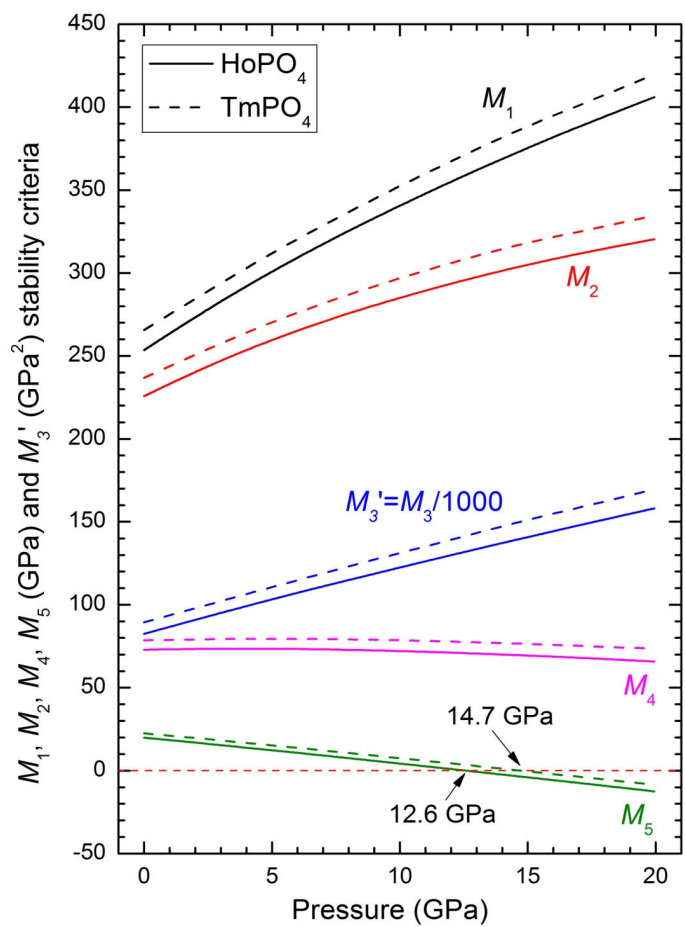

**Figure 11**

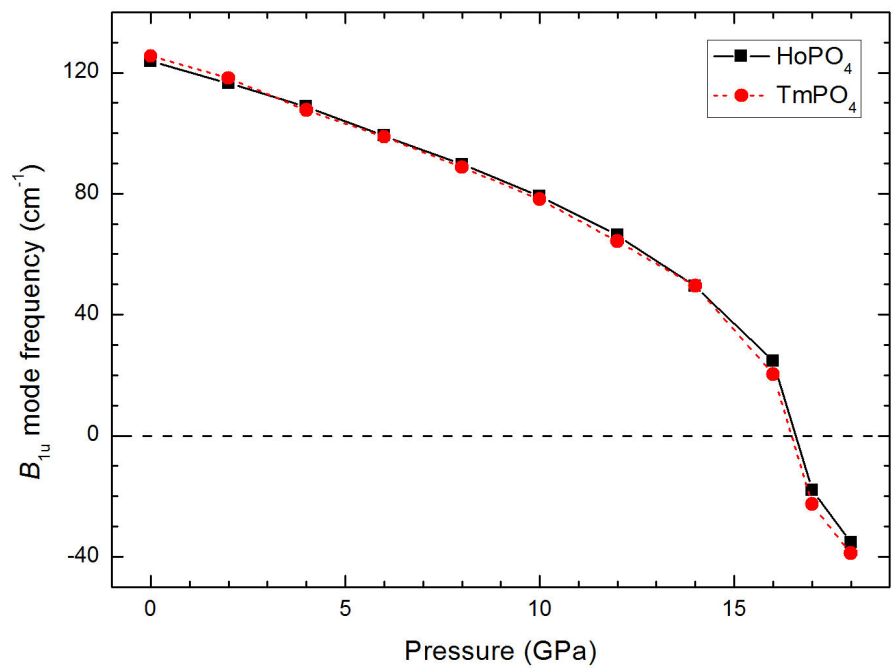